\documentclass[aps,prb,reprint,superscriptaddress,showpacs]{revtex4-2}
\usepackage[dvipdfmx]{graphicx}
\usepackage{amsmath}
\usepackage{float}
\usepackage{subfigure}
\usepackage{tikz}
\usepackage[english]{babel}
\usepackage{amssymb}
\usepackage{mathrsfs}
\usepackage{physics}
\usepackage[utf8]{inputenc}
\usepackage{booktabs}
\usepackage{hyperref}
\usepackage[normalem]{ulem}
\usepackage{xcolor}

\newcommand{\non}{\nonumber}

\def\Hex{H_{\mathrm{ex}}}

\begin{document}

\title{Probing valley quantum oscillations via the spin Seebeck effect in transition metal dichalcogenide/ferromagnet hybrids}

\author{Xin Hu }
\affiliation{%
Kavli Institute for Theoretical Sciences, University of Chinese Academy of Sciences, Beijing, 100190, China.
}%

\author{Yuya Ominato}
\affiliation{%
Waseda Institute for Advanced Study, Waseda University, Shinjuku-ku, Tokyo 169-0051, Japan.
}%

\author{Mamoru Matsuo }
\email{mamoru@ucas.ac.cn}
\affiliation{%
Kavli Institute for Theoretical Sciences, University of Chinese Academy of Sciences, Beijing, 100190, China.
}%
\affiliation{%
CAS Center for Excellence in Topological Quantum Computation, University of Chinese Academy of Sciences, Beijing 100190, China
}%
\affiliation{%
Advanced Science Research Center, Japan Atomic Energy Agency, Tokai, 319-1195, Japan
}%
\affiliation{%
RIKEN Center for Emergent Matter Science (CEMS), Wako, Saitama 351-0198, Japan
}%

\date{\today}

\begin{abstract}
 We theoretically investigate spin-valley-locked tunneling transport in a transition-metal dichalcogenide/ferromagnetic-insulator heterostructure under a perpendicular magnetic field, driven by the spin Seebeck effect. We demonstrate that spin-valley coupling together with the magnetic-field-induced valley-asymmetric Landau-level structure enables the generation of a valley-polarized spin current from valley-selective spin excitation. We compare the spin current and the valley-polarized spin current in the conduction and valence bands and clarify their distinct microscopic origins. We predict pronounced quantum oscillations of the valley-polarized spin current, providing a clear experimental signature of quantized valley states.
\end{abstract}

  \maketitle

\section{Introduction}
Monolayer transition-metal dichalcogenides (TMDCs) have attracted significant attention in recent years as one of the best-known atomically thin semiconductor platforms \cite{kormanyos2015k,liu2015electronic, xu2014spin} and as highly promising materials for future spintronic and valleytronic devices, where both the electron spin and valley are good quantum numbers and can serve as 
carriers of information and energy \cite{RevModPhys.76.323, PhysRevLett.97.186404, ye2016electrical}. 
Complementary to graphene, their strong intrinsic spin-orbit coupling (SOC) gives access to a variety of exotic magnetic phenomena \cite{sundararajan2025toward, abramchuk2018controlling, tartaglia2020accessing} and, together with broken inversion symmetry, gives rise to spin-valley coupling (SVC) \cite{PhysRevLett.99.236809,PhysRevB.77.235406,RevModPhys.82.1959,PhysRevB.81.195431,PhysRevB.84.125427}. 
This unique electronic structure allows the spin and valley degrees of freedom to be controlled individually by external fields, such as optical excitation via valley-selective circular dichroism \cite{caoValleyselectiveCircularDichroism2012, mak2012control, PhysRevB.86.081301} and electrical generation through the valley Hall effect or spin-polarized charge injection \cite{sanchezValleyPolarizationSpin2016, yeElectricalGenerationControl2016}, while their interplay \cite{zhuGiantSpinorbitinducedSpin2011,xiaoCoupledSpinValley2012,wang2015strong,zihlmann2018large, wakamura2018strong} provides a route to manipulating the valley polarization via pure spin injection.

\begin{figure}[htbp]
    \centering
    \includegraphics[width=0.5\textwidth]{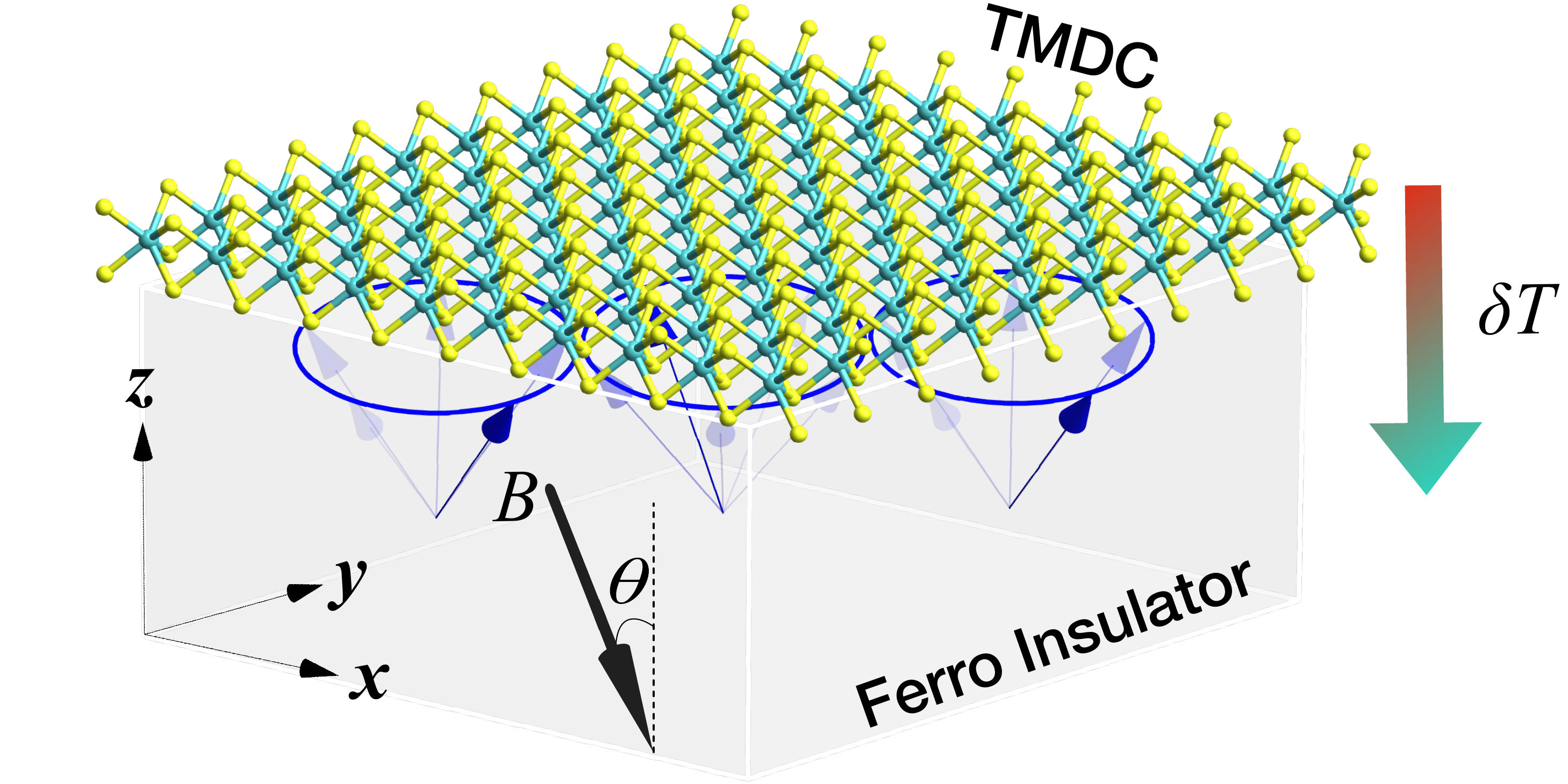}
    \caption{Schematic of the TMDC/ferromagnetic insulator heterostructure. A vertical temperature gradient $\delta T$ across the interface drives valley-polarized spin-tunneling transport via the spin Seebeck effect. An external magnetic field $\mathbf{B}$ is applied at a tilt angle $\theta$ relative to the surface normal ($z$-axis).}
    \label{system}
\end{figure}

To exploit this spin-valley coupling for valley transport, several theoretical works have proposed using TMDC/ferromagnetic-insulator (FI) hybrids, where magnon-driven spin transport at a magnetic interface plays a central role \cite{ominato_valley-dependent_2020,hu2025quantumoscillationsvalleycurrent}. In these systems, spin pumping (SP) provides a mechanism by which microwave irradiation excites magnons in the FI layer, and the interfacial exchange coupling transfers angular momentum from these magnons to the electron spins in the TMDC layer, thereby generating spin-valley-locked tunnel transport. 
Spin pumping has long been a key method for injecting spin currents into a wide variety of materials in spintronics \cite{mizukami2001study,mizukami2001ferromagnetic,mizukami2002effect,tserkovnyak2002enhanced,vzutic2004spintronics,tserkovnyak2005nonlocal,Hellman2017-fm}. 
 A closely related mechanism is the spin Seebeck effect (SSE) \cite{Uchida2008-we,xiao2010,adachi2011,bauer2012spin}, in which a temperature gradient replaces microwave excitation and drives tunnel spin transport across magnetic interfaces. The SSE has already been demonstrated in many types of heterostructures~\cite{Uchida2010-hr,jaworski2010observation,uchida2010spin,Kikkawa2023-va} and, when combined with the intrinsic valley selectivity of TMDCs, offers a promising route to drive spin-valley transport in TMDC/FI hybrids and to further exploit the multifunctionality of TMDC-based spintronic and valleytronic devices.

In this article, we theoretically investigate thermally driven spin and valley transport in a transition-metal dichalcogenide (TMDC)/ferromagnetic insulator (FI) heterostructure under a perpendicular magnetic field, as illustrated in Fig.~\ref{system}. An interfacial temperature gradient excites magnons in the FI, which in turn drive spin-flip processes in the TMDC via the spin Seebeck effect (SSE). The paper is organized as follows: In Sec.~\ref{system Hamiltonian}, we introduce the effective Hamiltonian for the hybrid system. Section~\ref{Valley} formulates the theoretical framework, demonstrating how the valley-selective spin excitation arises from the interplay of spin-valley coupling and the valley-asymmetric Landau level structure. In Sec.~\ref{numerical}, we present numerical results for both the conduction and valence bands, clarifying the microscopic origins of the valley-polarized spin current and its pronounced quantum oscillations. We propose an experimental detection scheme in Sec.~\ref{Experiment} and discuss the experimental feasibility and advantages of our thermal approach compared to conventional methods in Sec.~\ref{Discussion}. Finally, we summarize our findings in Sec.~\ref{Conclusion}. Our work suggests that the SSE provides a robust, all-spin-based pathway to manipulate valley degrees of freedom, advancing the integration of spintronics and valleytronics.

\section{System Hamiltonian}\label{system Hamiltonian}

The heterostructure consists of a monolayer TMDC placed on an FI substrate, as illustrated in Fig.~\ref{system}. The system is subjected to a perpendicular magnetic field $B$ and a thermal gradient $\delta T$ parallel to the magnetic field. The external magnetic field is tilted by an angle $\theta$ from the perpendicular direction, which is important to detect the SSE (see Sec.~\ref{Experiment}). In the following calculations, we set $\theta=0$ . The results for $\theta \neq 0$ are obtained by replacing $B$ with $B \cos \theta $ .
The total Hamiltonian is expressed as the sum of three components:
\begin{align}
    H = H_{\mathrm{TM}} + H_{\mathrm{FI}} + H_{\mathrm{ex}}.
\end{align}
Here, $H_{\mathrm{TM}}$ describes the electrons in the TMDC, $H_{\mathrm{FI}}$ represents the magnons in the FI, and $H_{\mathrm{ex}}$ accounts for the exchange interaction at the TMDC/FI interface.

The low-energy electronic states in the TMDC near the $K$ ($\tau=+$) and $K'$ ($\tau=-$) valleys are well captured by the effective Hamiltonian~\cite{xiaoCoupledSpinValley2012}:
\begin{align}\label{eq_TMDC_eff}
    H_{\mathrm{eff}} = v\left( \tau \pi_{x} \sigma^{x} + \pi_{y} \sigma^{y}\right) + \frac{\Delta}{2} \sigma^{z} - \tau s\lambda \frac{\sigma^{z} - 1}{2},
\end{align}
where $\vb*{\pi} = \vb*{p} + e\vb*{A}$ is the kinetic momentum with the vector potential $\vb*{A}$ in the Landau gauge, $v$ is the velocity, $\sigma^{x,y,z}$ are the Pauli matrices acting on the sublattice space, $\Delta$ is the energy gap, and $\lambda$ is the spin splitting at the valence-band top caused by spin-orbit coupling. The index $s=\pm$ denotes the electron spin. These parameters are obtained by fitting to first-principles calculations~\cite{xiaoCoupledSpinValley2012,zhuGiantSpinorbitinducedSpin2011,cheiwchanchamnangijQuasiparticleBandStructure2012, echeverrySplittingBrightDark2016, kormanyos2015k, kormanyosMonolayerMoS22013a, liuThreebandTightbindingModel2013}.

The Hamiltonian for the FI, within the spin-wave approximation, describes magnons in the bulk FI:
\begin{align}
    H_{\mathrm{FI}}(t)\simeq\sum_{\vb*{k}} \hbar \omega_{\vb*{k}} b_{\vb*{k}}^{\dagger} b_{\vb*{k}},
\end{align}
where $b_{\vb*{k}}^{\dagger}$ ($b_{\vb*{k}}$) is the creation (annihilation) operator for a magnon with wave vector $\vb*{k}$ and energy $\hbar\omega_{\vb*{k}}$ given by the Holstein-Primakoff transformation \cite{PhysRev.58.1098} and employing the spin-wave approximation: $S_{\vb*{k}}^+\simeq\sqrt{2S}b_{\vb*{k}}$, where $S_{\vb*{k}}^{\pm}$ are the Fourier components of the ladder operators of the localized spin in the FI and $S$ is the amplitude of the spin per site. In the following calculations, we assume that $\hbar\omega_{\vb*{k}}$ is given by 
\begin{align}\label{DIS}
    \hbar\omega_{\vb*{k}}=2\mathcal{J}Sa^2k^2+\hbar\gamma B ,
\end{align}
where $a$ is the lattice constant of the FI, $\mathcal{J}$ is the exchange coupling between nearest-neighbor spins in the FI, and $\gamma$ is the gyromagnetic ratio.

The exchange interaction at the interface, $\Hex$, is decomposed into a Zeeman-like term $H_Z$ \cite{PhysRevB.92.121403, zhang2016large, liang2017magnetic, PhysRevB.97.041405, PhysRevB.96.085411, PhysRevLett.122.086401} and a tunneling term $H_T$~\cite{kato_microscopic_2019,PhysRevLett.120.037201, ohnumaTheorySpinPeltier2017, ohnuma_enhanced_2014,PhysRevLett.122.086401}:
\begin{align}
    &\Hex
    =-\int d\vb*{r}\sum_j J(\vb*{r},\vb*{r}_j)\vb*{s}(\vb*{r})\cdot\vb*{S}_j
    =H_Z+H_T, \\
    &H_Z=-\int d\vb*{r}\sum_jJ(\vb*{r},\vb*{r}_j)s^z(\vb*{r})S^z_j \simeq -J_0S_0 s^z_{\mathrm{tot}}, \\
    &H_T=-\frac{1}{2}\int d\vb*{r}\sum_j
    J(\vb*{r},\vb*{r}_j)
    \qty(s^+(\vb*{r})S^-_j+s^-(\vb*{r})S^+_j),
\end{align}
where $s^z_{\mathrm{tot}}$ is the $z$-component of the total electron spin, $s^\pm(\vb*{r})$ are the ladder operators in the TMDC and $\vb*{S}_{j}$ is the localized spin in the FI at site $j$. 
The term $H_Z$ arises from the perpendicular component of the exchange coupling, inducing a proximity-exchange field, while $H_T$ describes the spin-flip processes between TMDC electrons and FI magnons.

We treat $H_T$ as a perturbation of the unperturbed Hamiltonian $H_0 = H_{\mathrm{TM}} + H_{\mathrm{FI}} + H_Z$. The Zeeman-like term $H_Z$ produces a spin-dependent shift of the electronic levels, $-sJ_0S_0$. Diagonalizing $H_{\mathrm{eff}} + H_Z$ in a magnetic field yields the Landau level spectrum:
\begin{align}
\varepsilon_{n,\tau,s}=\varepsilon_0\operatorname{sgn}_{\tau}(n) \sqrt{|n|+(\Delta')^2}+\frac{\tau s \lambda}{2}-sJ_0S_0,
\end{align}
where $\varepsilon_0 = \sqrt{2}\hbar v/l_B$ with the magnetic length $l_B=\sqrt{\hbar/eB}$, and $\Delta' = (\Delta/2 - \tau s \lambda/2)/\varepsilon_0$. The function $\mathrm{sgn}_\tau(n)$ is defined as $+1$ for $n>0$, $-1$ for $n<0$, and $\tau$ for $n=0$. 

\begin{figure}[htbp]
    \centering
    \includegraphics[width=0.49\textwidth]{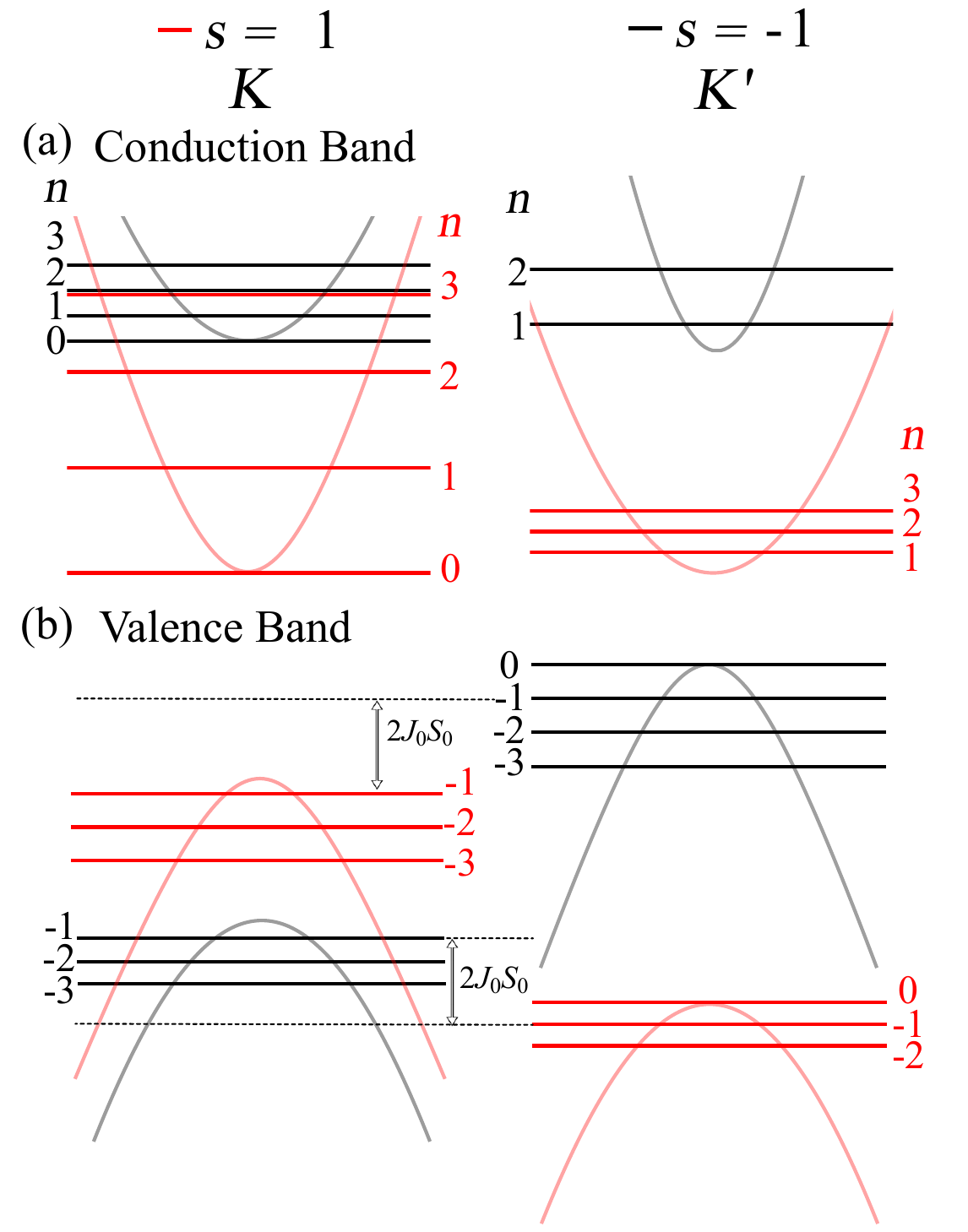}
    \caption{The energy spectrum of the TMDC. The curves represent the energy bands without a magnetic field, and the lines represent the Landau levels in the presence of a perpendicular magnetic field. }
    \label{LL}
\end{figure}

The Landau level structure is illustrated in Fig.~\ref{LL}. In each electron band, the valley degeneracy is lifted. The asymmetry between the two valleys is mainly manifested in two respects: For the conduction band, as shown in subfigure (a), the $n=0$ Landau level exists only in the $K$ valley, whereas in the valence band, as shown in subfigure (b), the $n=0$ Landau level exists only in the $K'$ valley. Moreover, in the valence band, the exchange splitting $J_0S_0$ causes the spin-up Landau levels to move up by $J_0S_0$ and the spin-down ones to move down by the same amount in both valleys.
Such a valley-asymmetric Landau level structure provides the key mechanism that enables the generation of a valley-polarized spin current.

\section{Spin current for each valley}\label{Valley}

\begin{figure*}[htbp]
    \centering
    \includegraphics[width=01.0\textwidth]{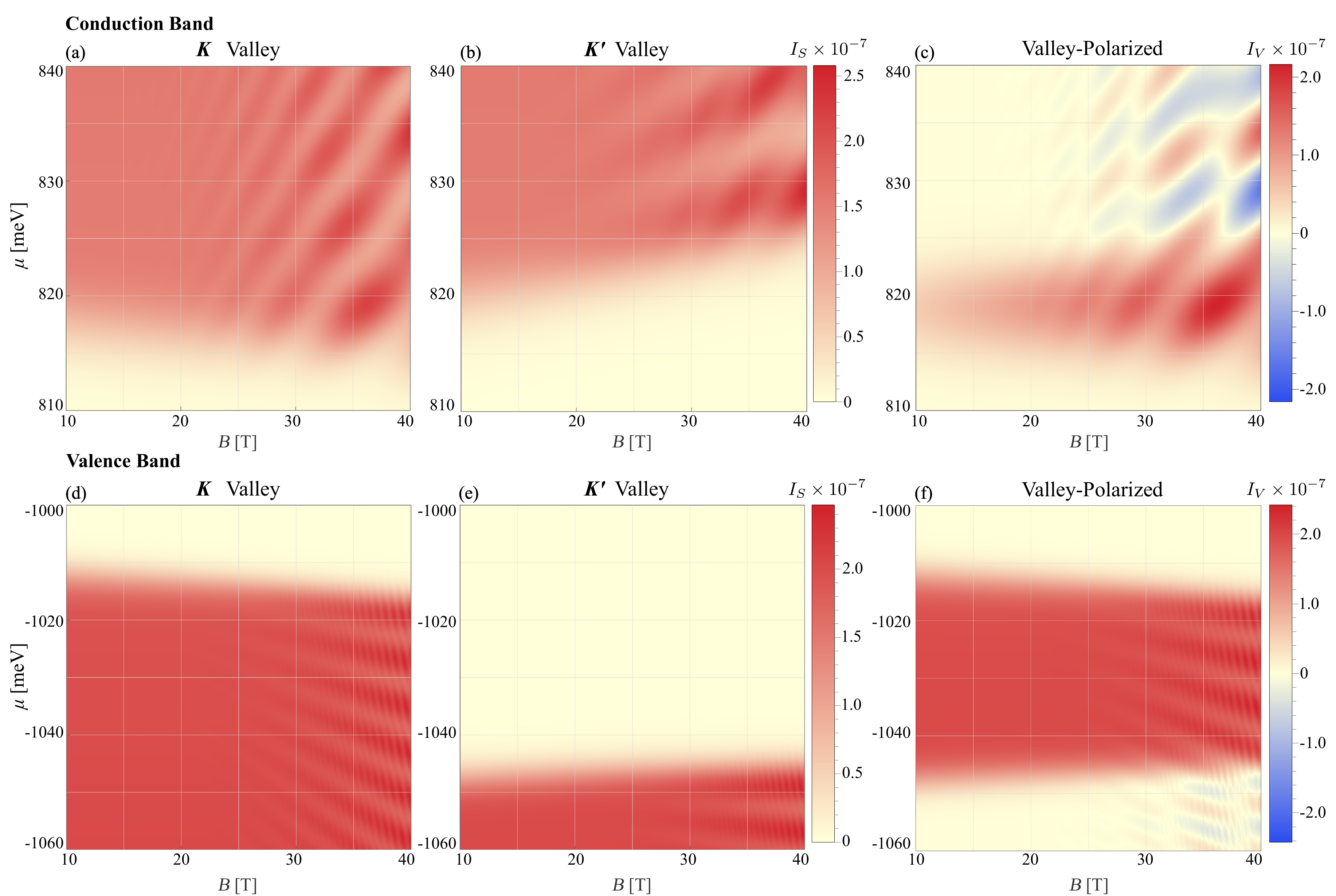}
    \caption{ The density plots of the dimensionless spin current generated by the SSE for the $K$ and $K'$ valleys in the conduction band [subfigures (a)-(b)] and the valence band [subfigures (d)-(e)] and the corresponding valley-polarized spin current in the conduction band [subfigure (c)] and the valence band [subfigure (f)] as functions of the chemical potential $\mu$ and magnetic field $B$. Simulation parameters: $k_BT=1$ meV, $J_0S_0=20$ meV, and $\Gamma=2$ meV.}
    \label{density}
\end{figure*}

The temperature gradient across the TMDC/FI interface generates a tunneling spin current through spin-flip processes.
The operator for the $z$-component of the spin current $I_s^z$ is defined as the rate of change of the total electron spin in the TMDC with respect to time: $I_s^z:=-\frac{\hbar}{2}\dot{s}^z_{\mathrm{tot}}$.

We calculate the statistical average of the spin current within second-order perturbation theory. Accordingly, the spin current is given by \cite{ohnuma_theory_2017,hu_spin_2024,ominato_valley-dependent_2020,kato_microscopic_2019, Ominato2022-xc, Ominato2022-se,m_matsuo_spin_2018, ohnuma_enhanced_2014}:
\begin{align} 
    \langle I_{s}^z (t) \rangle=I_{s,\tau=1} + I_{s,\tau=-1},
\end{align}
where the total spin current is a sum of contributions from the $K (\tau=1)$ and $K' (\tau=-1)$ valleys. In each valley component $I_{s,\tau}$, the valley-dependent spin current is given by:
\begin{align}\label{valley_spin_current}
I_{s,\tau}=
    \hbar J_0^2A\int \frac{d\omega}{2\pi}
    &\mathrm{Im}\chi^R_{\tau,\mathrm{loc}}(\omega)
    \qty[-\mathrm{Im}G^R_{\mathrm{loc}}(\omega)] \non \\
    & \times\qty(
        \frac{\partial f_{\mathrm{BE}}(\omega,T)}{\partial T} 
    )\delta T,
\end{align}
where $f_{\mathrm{BE}}(\varepsilon)$ is the Bose distribution function, $A$ is the interface area and $J_0^2$ represents the configurationally averaged exchange coupling~\cite{hu_spin_2024, Ominato2022-se, Ominato2022-xc,kato_microscopic_2019}.

The local dynamic spin susceptibility of the TMDC electrons $\chi^R_{\mathrm{loc}}(\omega)$ is the first key component in the spin current, which is defined as:
\begin{align}
    \chi^R_{\mathrm{loc}}(\omega):=\int dt e^{i(\omega+i0)t}\frac{i}{\hbar}\theta(t)\langle[s^+(\vb*{0},t),s^-(\vb*{0},0)]\rangle_0 ,
\end{align}
where the average $\langle \cdots \rangle_0$ is taken over the unperturbed Hamiltonian. Its imaginary part, which quantifies the density of available spin-flip excitations, is also a combination of contributions from both valleys, $\mathrm{Im}\chi^R_{\mathrm{loc}}(\omega)=\sum_{\tau}\mathrm{Im}\chi^R_{\tau,\mathrm{loc}}(\omega)$. The valley-decomposed spin susceptibility is calculated as:
\begin{align}\label{chi}
    \operatorname{Im}\chi _{\tau,\mathrm{loc}}^{R}( \omega )=
    \pi \int d\varepsilon
    &  D^\mathrm{LL}_{\tau,+1}(\varepsilon)D^\mathrm{LL}_{\tau,-1}(\varepsilon+\hbar\omega)W(\varepsilon,\tau) \non \\
    & \times  [ f_{\mathrm{FD}}(\varepsilon)-f_{\mathrm{FD}}(\varepsilon+\hbar\omega)] ,
\end{align}
where $f_{\mathrm{FD}}(\varepsilon)$ is the Fermi-Dirac distribution function. The density of states (DOS) for the Landau levels, $D^\mathrm{LL}_{\tau,s}(\varepsilon)$, is defined using a Gaussian function to account for level broadening $\Gamma$~\cite{doi:10.1021/ed044p432,GaussianLorentzianSumProduct2018}:
\begin{align}
    D_{\tau ,s}^\mathrm{LL}( \varepsilon ) = \frac{1}{2\pi l_{B}^{2}}\sum _{n}\frac{1}{\Gamma\sqrt{\pi} }  \exp[-\frac{(\varepsilon -\varepsilon _{n,\tau ,s})^2}{2\Gamma^2}].
\end{align}
The function $W(\varepsilon,\tau)$ represents the matrix element for spin-flip transitions, which depends on the orbital character of the wave functions:
\begin{align}
    W( \varepsilon ,\tau ) =\frac{1}{2} +\frac{1}{8}\frac{\Delta ^{2} -\lambda ^{2}}{\left( \varepsilon -\frac{\tau \lambda }{2} +J_{0} S_{0}\right)\left( \varepsilon +\frac{\tau \lambda }{2} -J_{0} S_{0}\right)},
\end{align}
which stems from $|\phi_{p}^{\dagger}\phi_q|^2$ as the overlap of the eigenfunction of Eq.~(\ref{eq_TMDC_eff}).

The magnon propagator is defined as $ G^R(\vb*{k},\omega):=\int dte^{i(\omega+i0)t}(2 S / i \hbar) \theta(t)\langle[b_{\vb*{k}}(t),b_{\vb*{k}}^{\dagger}(0)]\rangle_0$ and calculated as:
\begin{align}
    G^{R}(\vb*{k},\omega)
    =\frac{2S}{\hbar}\frac{1}{\omega-\omega_{\vb*{k}}+i0}.
\end{align}
The density of states of magnons per site is represented by the magnon propagator:
\begin{align}
    D_m(\varepsilon)=-\frac{1}{2\pi S}\mathrm{Im}G^R_{\mathrm{loc}}(\varepsilon/\hbar),
\end{align}
where $G^R_{\mathrm{loc}}(\omega):=\frac{1}{N}\sum_{\vb*{k}}G^R(\vb*{k},\omega)$ is the local magnon propagator in the FI.

For simplicity, we define the weight function as the product of the DOS function of magnons and the derivative of the Bose distribution function:
\begin{align}
    M(\omega,T)=\frac{D_m(\omega)}{(\mathcal{J}S)^{-3/2}} \frac{\partial f_{\mathrm{BE} }}{\partial T}, 
\end{align}
where we have normalized the weight function by $(\mathcal{J}S)^{-3/2}$ to make it dimensionless. 
Thus, the valley-dependent spin current in Eq.~(\ref{valley_spin_current}) is rewritten as:
\begin{align}
 I_{s,\tau}
    =\frac{\pi SJ_0^2 A k_{\mathrm{B}}\delta T }{(\mathcal{J}S)^{3/2}}\int _{0 }^{E_M }  d(\hbar \omega) \mathrm{Im}\chi^R_{\tau,\mathrm{loc}}(\omega) M(\hbar \omega,T).
\end{align} 
where $E_M$ is the high-energy cutoff of the magnon dispersion relation.

\begin{figure*}[htbp]
    \centering
    \includegraphics[width=01.0\textwidth]{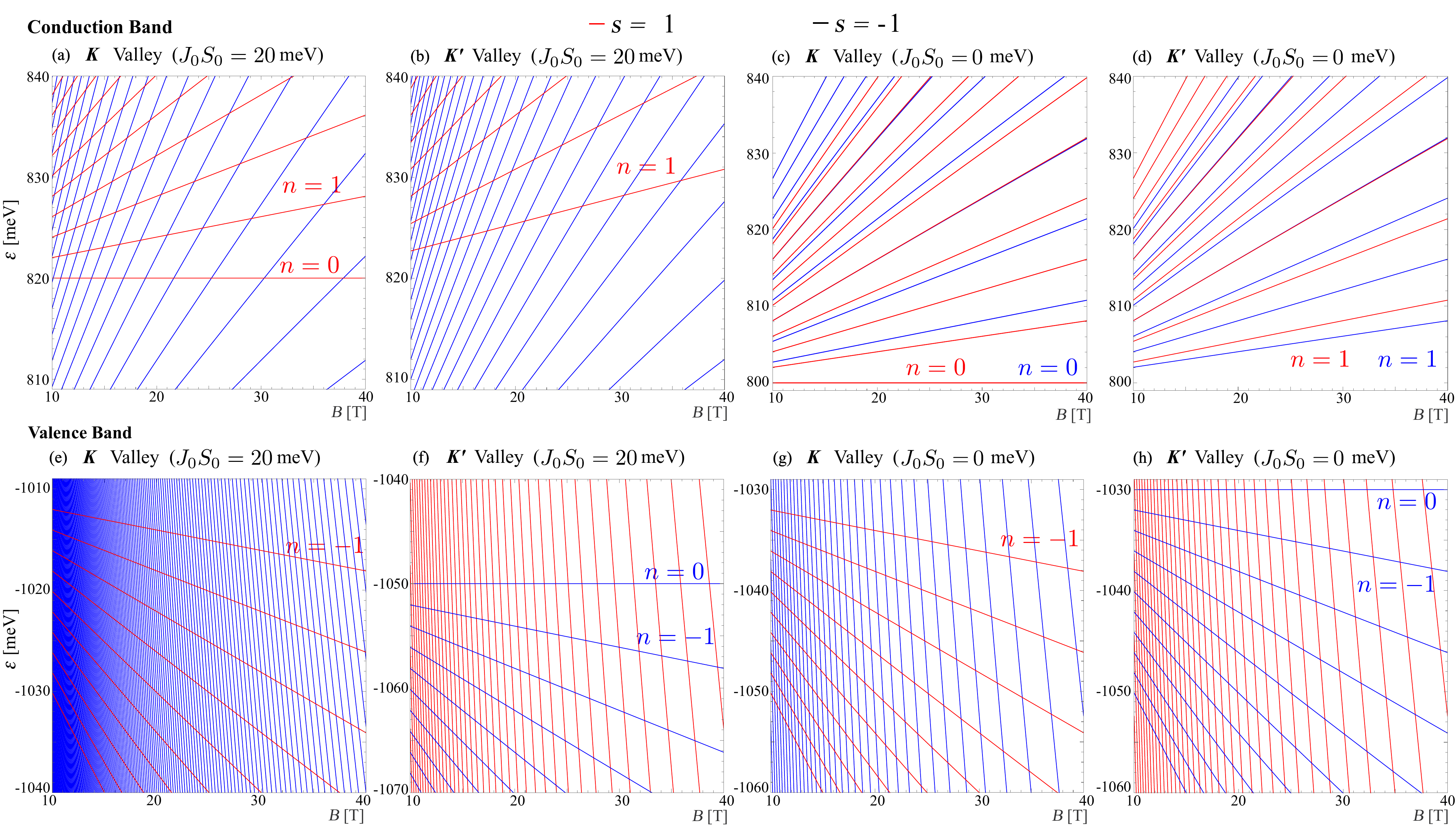}
    \caption{ The Landau level (LL) spectrum in both the conduction band [subfigures (a)-(d)] and the valence band [subfigures (e)-(h)] for both valleys and under two different spin-splitting strengths $J_0S_0=20$ meV and $J_0S_0=0$ meV. 
    For the conduction band: (a) the spin-up LL indices are $n=0,1,2,\cdots $ and in (b) the spin-up LL indices are $n=1,2,\cdots $ with the absence of the $n=0$ LL. When the spin-splitting strength is $J_0S_0=0$ meV in (c) both the spin-up and spin-down LL indices are $n=0,1,2,\cdots$ and in (d) both the spin-up and spin-down LL indices are $n=1,2,\cdots $ with the absence of the $n=0$ LL.
    For the valence band: (e) the spin-up indices are $n=-1,-2,-3,\cdots $ and in (f) the spin-down LL indices are $n=0,-1,-2,-3,\cdots $. When the spin-splitting strength is $J_0S_0=0$ meV, in (g) the spin-up indices are $n=-1,-2,-3,\cdots $ and in (h) the spin-down LL indices are $n=0,-1,-2,\cdots $.
    Except for $J_0S_0$, other simulation parameters are the same as Fig.~\ref{density}.}
    \label{LL_schematic}
\end{figure*}

\section{Generation of valley-polarized spin current and the numerical results}\label{numerical}
The valley-dependence of the spin current for each valley indicates that magnon-mediated spin excitation is valley-selective. This enables the generation of a valley-polarized spin current, defined as the difference between the two spin current components for each valley:
\begin{align}
    I_v^z=I_{s,K} - I_{s,K'}.
\end{align}

In the following analysis, all currents are normalized by
\begin{align}
    I_0=
    \frac{\pi SJ_0^2 A k_{\mathrm{B}}\delta T }{(\mathcal{J}S)^{3/2}}\frac{1}{[\mathrm{meV}]^2[\mathrm{nm}]^4}.
\end{align}
In the following numerical simulations, we use the parameters of WSe$_2$ \cite{xiaoCoupledSpinValley2012,zhuGiantSpinorbitinducedSpin2011} as a representative TMDC material: $v=5.3\times10^5$ m/s, $\Delta=1.6$ eV, and $\lambda=230$ meV.

Figure~\ref{density} presents the density plots of the spin current and the corresponding valley-polarized spin current ($I_v$) for both the conduction and valence bands of the TMDC/FI heterostructure as functions of chemical potential $\mu$ and external magnetic field $B$. A prominent $I_v$ is observed in both bands, stemming from the disparity between the spin currents in the $K$ and $K'$ valleys. Furthermore, distinct striations appear in the density plots for both the valley-dependent spin currents and the total valley-polarized spin currents $I_v$, signifying quantum oscillations derived from the discrete Landau level (LL) spectrum. These oscillations are evident in the line plots in Figs.~\ref{Con} and \ref{Val}. These oscillatory signals become more robust and clearly resolved with increasing magnetic field. 

The distinct transport behaviors in the conduction and valence bands can be attributed to the asymmetry in their respective Landau level structures. This asymmetry originates from two primary mechanisms: (i) the intrinsic valley-asymmetric appearance of the $n=0$ Landau level, and (ii) the extrinsic energy shift induced by the proximity exchange interaction $J_0S_0$. We discuss these mechanisms below, referencing the LL spectra shown in Fig.~\ref{LL_schematic}. The shapes of the bright ranges in Fig.~\ref{density} (a), (b), (d), and (e) essentially correspond to the LL spectrum structure in Fig.~\ref{LL_schematic}, indicating the pivotal role that the LL structure plays in the behavior of the spin currents.

\subsection{Conduction Band}

In the conduction band, the asymmetry is dominated by the intrinsic topological properties of the valleys. As shown in Figs.~\ref{LL_schematic}(a) and (b), the $K$ valley hosts a unique, field-independent $n=0$ Landau level at the band edge, whereas the spectrum in the $K'$ valley begins at the $n=1$ level, which shifts to higher energies with increasing $B$.

\begin{figure}[h]
    \centering
    \includegraphics[width=0.5\textwidth]{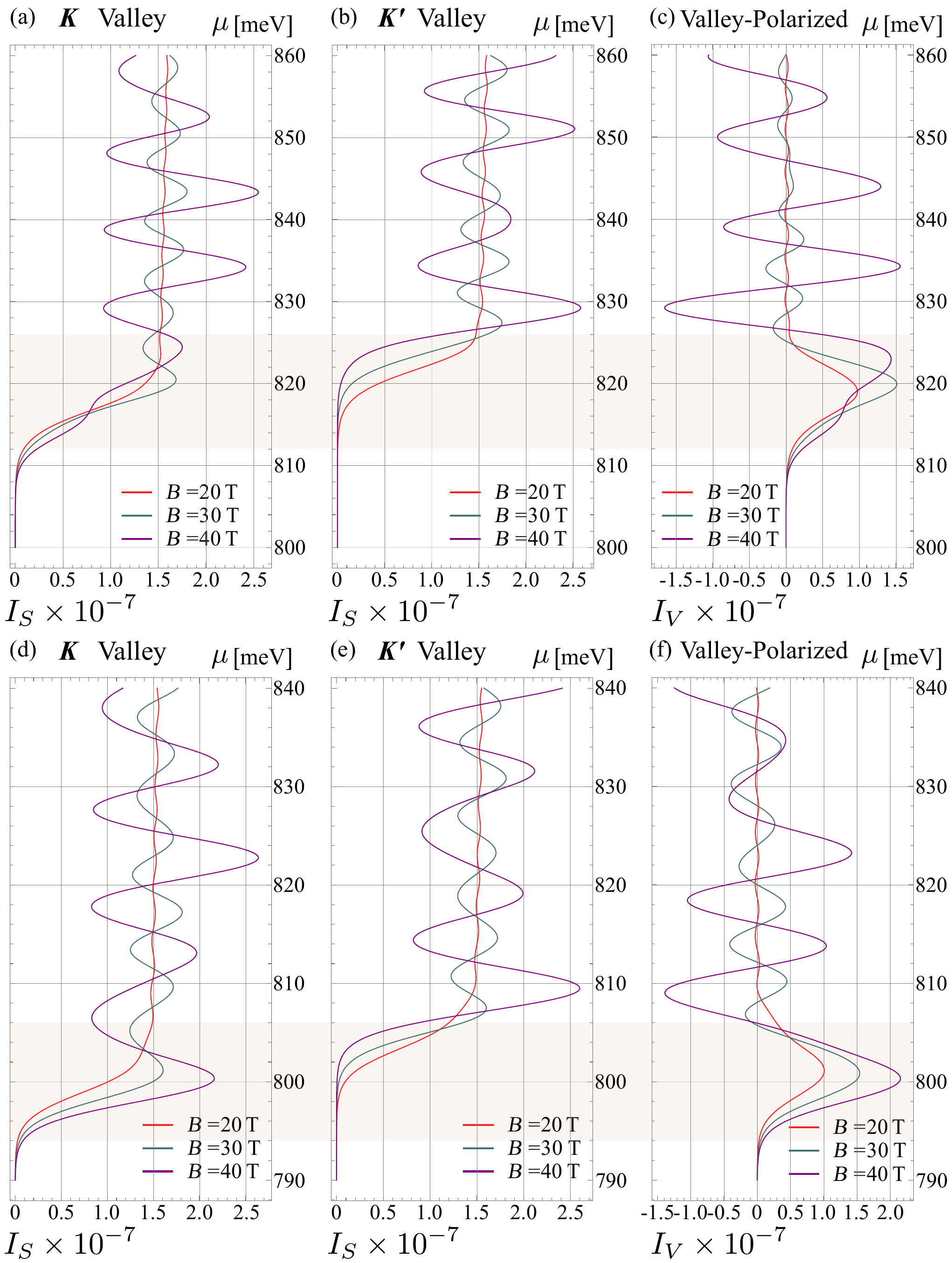}
    \caption{ The spin currents for each valley and the valley-polarized spin current 
    in the conduction band as a function of chemical potential $\mu$ for magnetic fields $B=20, 30, 40$ T. 
    (a) Spin current from the $K$ valley, $I_{s,K}$ when $J_0S_0=20$ meV. 
    (b) Spin current from the $K'$ valley, $I_{s,K'}$ when $J_0S_0=20$ meV.  
    (c) The resulting valley-polarized spin current, $I_v$ when $J_0S_0=20$ meV.  Orange shading highlights the range of significant valley polarization.
    (d)-(f) are the corresponding spin-current behavior when $J_0S_0=0$ meV. A detailed comparison is discussed in the text.
    Simulation parameters: $k_BT=1$ meV and $\Gamma=2$ meV.}
    \label{Con}
\end{figure}

This structural difference leads to the asymmetry observed in Figs.~\ref{density}(a) and (b) and the corresponding line plots in Figs.~\ref{Con}(a),(b),(d), and (e). For the $K$ valley, the stable $n=0$ level pins the lower boundary of the spin current existence range. In contrast, for the $K'$ valley, the boundary shifts to higher chemical potentials as the lowest available level ($n=1$) moves up with the magnetic field. Consequently, a finite window emerges where current exists primarily in one valley, resulting in the emergence of the valley-polarized spin current $I_v$ shown in Fig.~\ref{Con}(c) and (f), see also the triangular range of the valley-polarized spin current in Fig.~\ref{density}(c).

The influence of the proximity exchange splitting $J_0S_0$ is secondary in the conduction band. Since the asymmetry is fundamentally rooted in the presence of the $n=0$ level (which is robust against spin splitting), varying $J_0S_0$ merely shifts the energy levels without altering the order of the levels. 
Therefore, the profile of $I_v$ in the conduction band remains qualitatively similar regardless of the strength of $J_0S_0$. The highlighted orange region in Fig.~\ref{Con} marks the robust existence range of $I_v$ around the $n=0$ level energy. 

Notably, unlike the spin pumping (SP) scenario reported in Ref.~\cite{hu2025quantumoscillationsvalleycurrent}, the peak positions of the currents in this spin Seebeck effect (SSE) setup shift with the magnetic field. This is because the SSE involves a thermal broadening of the magnon energy $\hbar\omega$, contributing to spin-flip excitations over a wider range, whereas the SP scenario is restricted to $\omega \simeq 0$. Similar behavior has been predicted for graphene in SSE \cite{hu_spin_2024}.

\subsection{Valence Band}

In the valence band, the transport behavior is critically dependent on the strength of the proximity exchange splitting $J_0S_0$. Here, we define a large enough splitting as one for which $J_0S_0$ is significantly larger than the cyclotron energy gap between the zeroth and first Landau levels ($\Delta_{0,1} = |E_{n=1} - E_{n=0}|$).

As seen in Fig.~\ref{density}(d)-(f), a finite $I_v$ is also generated in the valence band. 
However, the sign and robustness of this current are determined by the interplay between the LL spacing and $J_0S_0$.
Figure~\ref{LL_schematic}(e) and (f) depict the LL spectrum for a large splitting ($J_0S_0=20$ meV). 
In this regime, the large $J_0S_0$ induces a significant energy gap between the spin-polarized sectors of the $K$ and $K'$ valleys since the order of the spin-up and spin-down series is inverse in each valley.
Specifically, the spin-up LL series in the $K$ valley and the spin-down LL series (now includes the $n=0$ level) in the $K'$ valley are energetically separated by about $2J_0S_0$. 

\begin{figure}[htbp]
    \centering
    \includegraphics[width=0.5\textwidth]{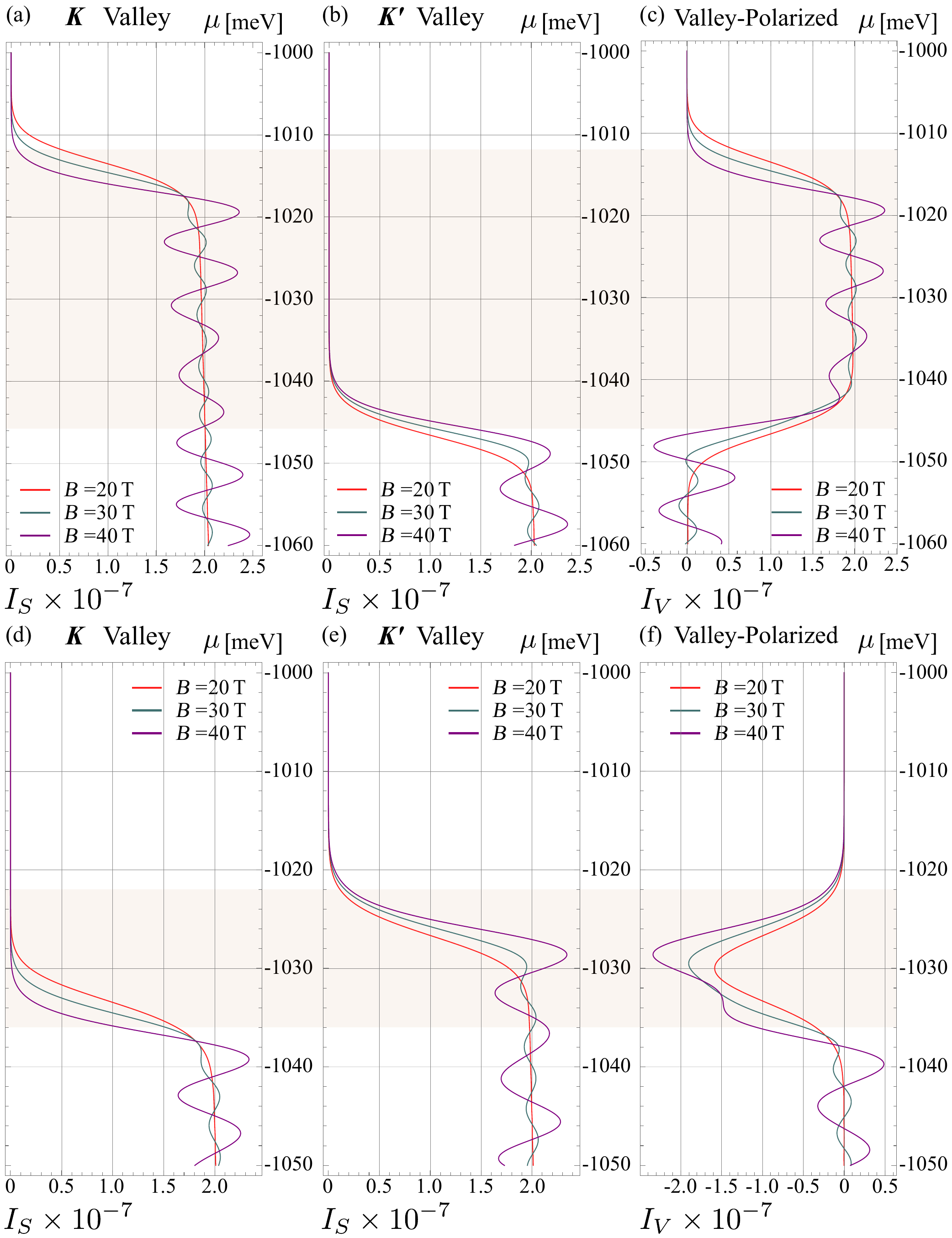}
    \caption{The spin currents for each valley and the valley-polarized spin current in the valence band. 
     Panels correspond to those in Fig.~\ref{Con}.
    (a)-(c) Case with large enough spin splitting ($J_0S_0=20$ meV). The valley-polarized spin current $I_v$ flows in the same direction as in the conduction band and exists over a wide range.
    (d)-(f) Case with zero spin splitting ($J_0S_0=0$ meV). The current flows in the opposite direction and is confined to narrow peaks.
    See Fig.~\ref{LL_schematic} for the corresponding LL structures.}
    \label{Val}
\end{figure}

The consequences of this splitting are illustrated in Fig.~\ref{Val}. When $J_0S_0$ is large enough (Fig.~\ref{Val}, top row), the induced gap ensures that the valley-polarized spin current $I_v$ flows in the positive direction following the sign of $J_0S_0$. The current is robust and spans a wide chemical potential range (indicated by orange shading), exhibiting quantum oscillations on the plateaus due to the dense distribution of high-index LLs.

Conversely, when $J_0S_0$ is small or zero (e.g., $J_0S_0=0$ meV), as shown in the schematic of Fig.~\ref{LL_schematic}(g)-(h), the LLs from different valleys overlap significantly. In this limit (Fig.~\ref{Val}, bottom row), the $I_v$ reverses direction relative to the conduction band and is only present as narrow peaks. In this situation, the mechanism of the emergence of the valley-polarized spin current is just the reversal of that in the conduction band, since now the $n=0$ level only exists in the $K'$ valley. 
Thus, the influence of the $n=0$ level and the spin-splitting $J_0S_0$ compete in the valence band, determining the direction and stability of the valley-polarized current.

Finally, let us turn to the broad plateau-like structure observed in Fig.~\ref{Val} (c), which represents the spin Seebeck counterpart to the valley-selective spin pumping predicted in our previous work~\cite{ominato_valley-dependent_2020}. 
In the zero-field limit discussed therein, a large proximity-induced spin splitting $J_0S_0$ generates a continuous, flat spin current plateau. In contrast, the present high-field regime discretizes the density of states into Landau levels, transforming the featureless plateau into an oscillation-modulated structure. This implies that the associated spin-current Hall effect is not constant but is accompanied by pronounced quantum oscillations, providing a distinctive experimental signature of the valley-coupled Landau quantization.

\section{Experimental Detection}\label{Experiment}

\begin{figure}[htbp]
    \centering
    \includegraphics[width=0.5\textwidth]{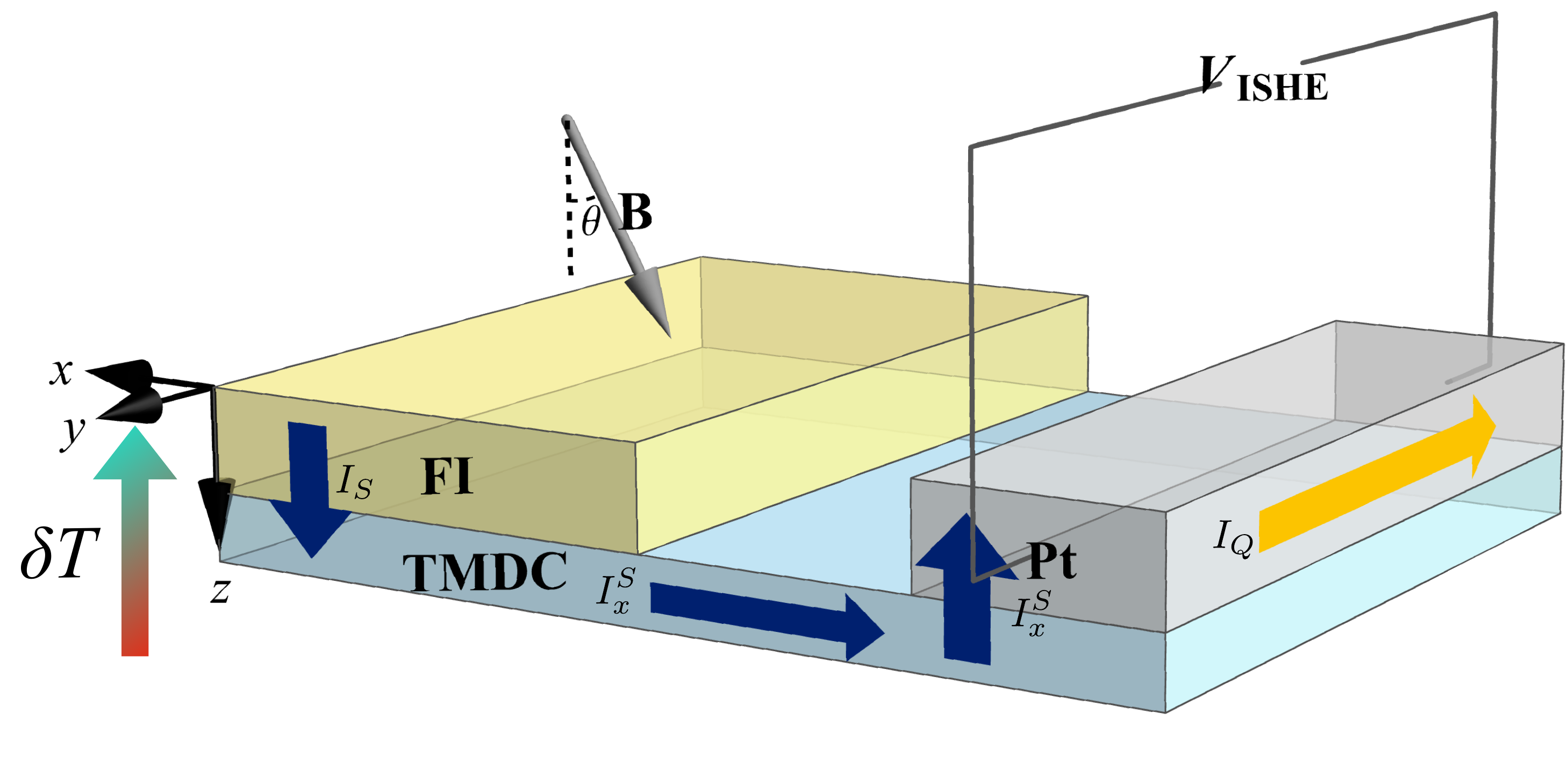}
    \caption{Schematic of the proposed experimental setup for the electrical detection of the spin current generated by the SSE. A vertical temperature gradient $\delta T$ drives spin injection from the FI into the TMDC layer. The external magnetic field $\mathbf{B}$ is tilted by an angle $\theta$ relative to the $z$-axis, inducing a finite in-plane spin polarization component (along the $x$-direction). These spins diffuse laterally through the TMDC channel and are injected into the Pt electrode, where the inverse spin Hall effect (ISHE) converts the spin current into a detectable transverse voltage $V_{\text{ISHE}}$ along the $y$-direction.}
    \label{ExpSet}
\end{figure}
We propose an experimental scheme for the electrical detection of the valley-polarized spin current generated by the SSE in the TMDC/FI heterostructure, as illustrated in Fig.~\ref{ExpSet}. The setup utilizes the inverse spin Hall effect (ISHE) in a heavy metal electrode (e.g., Pt) adjacent to the TMDC layer. 
In this configuration, a temperature gradient $\nabla T$ drives a spin current $I_s$ from the FI into the TMDC layer via the interfacial exchange coupling. To detect this current, the external magnetic field $\mathbf{B}$ is applied at a tilt angle $\theta$ relative to the surface normal ($z$-axis). This tilted geometry is crucial for the electrical detection of the SSE in lateral heterostructures. The spin polarization of the injected current is aligned with the external field direction. Consequently, the tilted field introduces a finite in-plane component of the spin polarization. As these spins diffuse laterally through the TMDC channel and are injected into the Pt electrode, the ISHE converts the spin current into a transverse charge current $I_Q$, resulting in a detectable voltage $V_{\text{ISHE}}$. 
The magnitude of the detected voltage is proportional to the projection of the spin polarization onto the plane of the film. Since both n- and p-type doping of TMDCs have been experimentally realized~\cite{HE2023107347,D2TC01045C}, the Fermi level can be tuned across the conduction and valence bands by applying a gate voltage $V_G$. This tunability allows for the mapping of the quantum oscillations predicted in our theory across different carrier regimes, providing a systematic verification of the valley-dependent transport.

\section{Discussion}\label{Discussion}
We now address the experimental feasibility and distinct advantages of the SSE method in the context of valleytronics. To resolve the pronounced quantum oscillations predicted in this work, the system requires a strong magnetic field (typically $B > 20$ T) to ensure that the Landau level spacing exceeds the thermal broadening. Such high-field environments have been experimentally realized in TMDC systems~\cite{ImagingQuantumSpin,kerdiHighMagneticField2020,liUnconventionalQuantumHall2013}. However, these conditions present a significant challenge for conventional spin pumping (SP) techniques. In SP, spin injection relies on ferromagnetic resonance (FMR), where the required microwave frequency scales linearly with magnetic field ($\omega = \gamma B$). At $B \approx 20$ T, the resonance frequency reaches the sub-terahertz regime, making resonant excitation technically demanding~\cite{hu2025quantumoscillationsvalleycurrent}. 
In contrast, the SSE is driven by incoherent thermal magnon excitations and does not require frequency matching. This feature renders the SSE particularly robust for exploring high-field phenomena where Landau quantization becomes dominant. Therefore, the SSE serves as a complementary approach to SP: while SP offers frequency-resolved spectroscopy primarily in lower field regimes, the SSE provides an effective route to access the high-field quantum Hall regime required to observe valley-coupled Landau quantization. The combination of these two methods would provide a useful toolkit for investigating spin-valley coupling in TMDC/magnet hybrids.

We finally contrast our mechanism with earlier proposals of valley-dependent Seebeck effects. Zhai et al. analyzed a valley-spin Seebeck effect in ferromagnetic heavy group IV monolayers, where a temperature gradient drives an electronic thermoelectric current in a single Dirac sheet whose valley and spin degrees of freedom are engineered by perpendicular electric fields and exchange fields \cite{zhaiValleySpinSeebeck2017}. Yu et al. proposed a valley Seebeck effect in gate-tunable zigzag graphene nanoribbons, in which a longitudinal temperature bias generates a pure valley-polarized spin current in a phase-coherent nanoribbon by exploiting momentum-valley locking and gate-controlled transmission asymmetry \cite{yuValleySeebeckEffect2016}. These works demonstrate that valley-polarized spin currents can in principle be generated by electronic thermoelectric transport, but they do not involve a magnetic insulator, magnon excitations, or Landau quantization. In contrast, our TMDC/ferromagnet hybrid realizes a magnonic spin Seebeck effect across a magnetic interface and uses the valley-spin-locked Landau level structure of the TMDC to convert the injected spin current into a quantum-oscillatory valley-polarized spin current. This hybrid, magnon-driven scheme provides a thermally driven probe of valley quantum oscillations at a magnetic interface, complementary to microwave-driven spin pumping and distinct from the purely electronic valley Seebeck scenarios discussed above.

\section{Conclusion}\label{Conclusion}
We theoretically investigate the spin Seebeck effect in a TMDC/FI heterostructure under a perpendicular magnetic field, focusing on the resulting spin current and the corresponding valley-polarized spin current. Our investigation reveals that spin--valley coupling, together with the magnetic-field-induced valley-asymmetric Landau level structure, enables the generation of a valley-polarized spin current from valley-selective spin excitation. We further clarify how the quantized and valley-asymmetric energy-level structure influences the spin current and the valley-polarized spin current in both the conduction and valence bands. In addition, we predict pronounced quantum oscillations of the valley-polarized spin current as a function of the chemical potential. These oscillations arise not only in the spin-pumping scenario but also in the thermally driven spin-injection setup studied here, providing a clear experimental signature for future experiments. These results provide a detailed understanding of the interplay between spintronics and valleytronics and offer a promising pathway for generating and controlling valley phenomena using pure spin excitation.

\begin{acknowledgments}
This work was supported by the National Natural Science Foundation of China (NSFC) under Grant No. 12374126, by the Priority Program of the Chinese Academy of Sciences under Grant No. XDB28000000, by JSPS KAKENHI under Grants (No. 21H01800, No. 21H04565, No. 23H01839, and No. 24H00322) from MEXT, Japan, and by Waseda University Grant for Special Research Projects (Grants No. 2025C-651 and No. 2025R-061).
\end{acknowledgments}


\bibliography{ref.bib}

\end{document}